\documentclass[twoside]{article}
\usepackage{fleqn,espcrc2}
\input{psfig}

\usepackage{graphicx}
\usepackage[figuresright]{rotating}

\newcommand{\AmS}{{\protect\the\textfont2
  A\kern-.1667em\lower.5ex\hbox{M}\kern-.125emS}}

\title{The Future of Diffraction at Tevatron}

\author{Alberto Santoro\address{Centro Brasileiro de Pesquisas F\'\i sicas, \\ 
        Rua Dr. Xavier Sigaud, 150 RJ - Rio de Janeiro, Brasil\\
        e-mail: santoro@cbpf.br}}
       
\begin{document}

\begin{abstract}
This paper summarizes the main current upgrade for Diffractive Physics at
Tevatron. We describe both CDF (Collider Detector Facility) and D\O\
new devices that are being installed around the CDF and D\O\ area that 
will allow to produce new results in Diffractive Physics. 
\vspace{1pc}
\end{abstract}

\maketitle

\section{Introduction}

Diffraction is certainly one of the oldest subject of physics. In Particle 
Physics we have seen a big development in the second half of this 
century~\cite{Predazzi}. More recently with the discovery of Hard Diffraction by 
UA8-Collaboration~\cite{UA8}, we have had a significant development of the 
Diffractive Physics from both theoretical and experimental point of view.
This is demonstrated in this workshop with many results coming from Hera (H1 and 
Zeus)~\cite{hera} in a large number of topics including diffractive structure
functions, and at the  Tevatron (CDF and D\O\ )  with results on 
diffractively produced jets at both energies $\sqrt{s}$ = 1.8 TeV 
and 630 GeV~\cite{santoro1}. From the theoretical side we had many presentations 
showing a great number of ideas about diffraction.  These results are important 
to understand the nature of the Pomeron. 
\par
	All the above experimental results constitute the data 
sample available for the development of theoretical and phenomenological models 
allowing the comparison with these predictions.   However many  
results have poor statistics and new data is needed. Some of the processes 
under study still have not been directly observed also.  Information about $t$ 
and $\xi$ 
distributions is missing to complete some studies and verify some 
theoretical results.  Despite of the theoretical progress for the 
unification of the soft and hard aspects of the strong interactions, we need to get
new experimental results to exploit deeply theoretical ideas. We have to stress 
again that 40\% of the cross section is due to the Pomeron exchange, and at higher 
energy can still be growing. This means that we have to pay 
attention to this physical region of the proton anti-proton interaction at the
Tevatron. We need more data to answer questions like: universality of the Pomeron, 
hadronic characteristic,
is the Pomeron a glueball?, is the Pomeron a dual object?, What is the real 
contribution for the cross sections of the diffractive heavy flavor production?,
Can Higgs and Centauros can be produced at Tevatron energies? What is the actual 
value of the $\overline{p} p$ total cross section? and so on.
\par
	All these questions lead to the proposal of new devices to 
be inserted at Tevatron, {\it Roman Pots Spectrometer + Miniplug +Beam Shower 
Counters} in the CDF area and a {\it Forward Proton Detector} (FPD) for the
D\O\ area. 
\par
\begin{figure}[htb]
\centerline{\psfig{figure=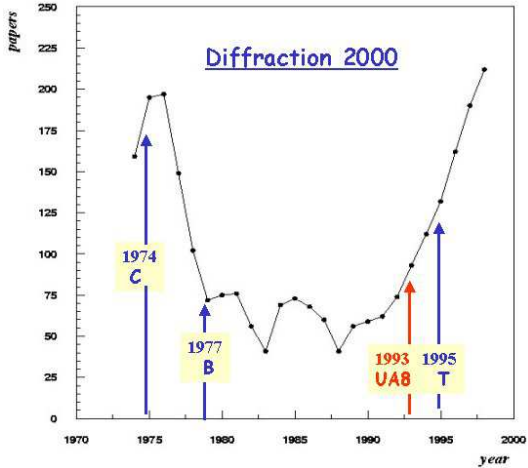,height=2.5in}}
\vspace{-1.cm}
\caption{Number of papers per year in diffractive physics~\cite{publicacoes}. We 
show four important dates which show some correlation with diffractive physics 
papers. 1974 was the year of charm discovery, three years later, in 1977 the 
bottom, in 1993 it was the year of the discovery of hard diffraction and 1995 the 
top was discovery.}
\label{fig:papers}
\end{figure}

	Figure~\ref{fig:papers} shows the growing interest for diffractive 
physics. There is a clear growing number of papers about this subject in 
particle physics.
\par	
	
	We expect soon a new period of diffractive results from 
several experiments. From the Tevatron, the next Run II with CDF and D\O\ ; from 
{\it  Brookhaven National Laboratory (BNL)}, {\it  Relativistic Heavy Ion 
Collider (RHIC)} in proton proton interactions at the energy of ${\sqrt{s}}$ = 50  
- 500 \ GeV; from the {\it Deutsches Elecktronen Synchroton (DESY) }, with H1 and 
ZEUS the two detectors at HERA in electron (30 GeV) proton (800 GeV) interactions. 
For the near
future we will have a new era of energy experiments, with TOTEM~\cite{giorgio} being
integrated to  {\it Compact Muon Solenoid (CMS)} as one of the  detectors of 
{\it Large Hadron Collider (LHC)} of {\it Centre European for Particle 
Physics (CERN)} for proton proton at 14 TeV energy in center of mass system.
\par 

\section{The Tevatron and the upgrades of CDF and D\O\ for Diffractive Physics}

	The Tevatron itself is being submitted to an upgrade in several parts of the
machine. The main parameters for physics are the energy going to ${\sqrt{s}}$ =
 2. \ TeV and the higher luminosity. The two detectors CDF and D\O\ are also
 being submitted to many upgrades. We will comment mainly those that have a 
direct relation with diffractive physics.
\par
	The Tevatron beam line was modified to open space for the FPD stations. The 
Electrostatic separator girder had to be modified, the cryogenic bypass was 
expanded to open space to allow the insertion of the Roman Pots
of FPD, the quadrupole magnet $Q_1$ is no longer present on the beam line. 
Other small modifications were necessary, like drilling a hole on the floor to 
allow the insertion and removal of the bottom detectors. 
	Summarizing the Tevatron modification are: (i) girder modification, (ii)
cryogenic bypass, (iii) removal of $Q_1$ quadrupole.
\par
	\subsection{CDF}
	We will summarize the upgrade for diffractive physics for CDF since the
reference~\cite{goulianos} shows a detailed description of this detector.
	The CDF upgrade for diffractive physics can be seen in three parts:
\begin{enumerate}
	\item{Improve dipole spectrometer (Roman Pots) as is shown in 
              figure~\ref{fig:cdfpots}}
\par	
\begin{figure}[htb]
\centerline{\psfig{figure=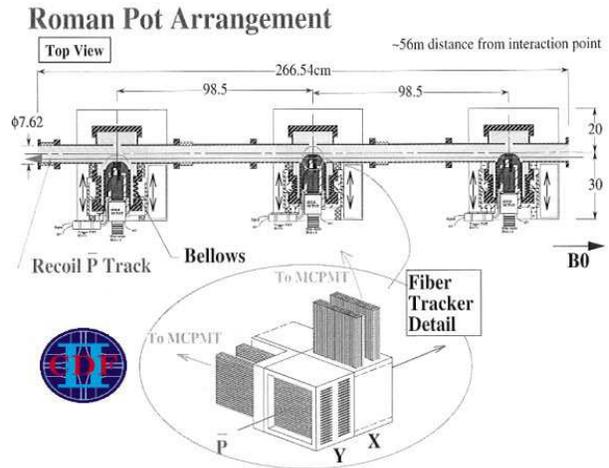,height=2.5in}}
\vspace{-1.cm}
\caption{Roman Pots arrangement of CDF.}
\label{fig:cdfpots}
\end{figure}		      
	\item{Use the miniplug for triggering Diffractive
              events. This device is shown in figure~\ref{fig:cdfminiplug}.
              Each Miniplug is composed by 50 lead plates of 1/4 thick 
              corresponding to a total of 2 interactions  lengths and ~60 
              radiation length. 288 signal towers viewed by 18 MCPMTs
              (16 channels each) and a total weight of  ~0.8 tons.}
\par	
\begin{figure}[htb]
\centerline{\psfig{figure=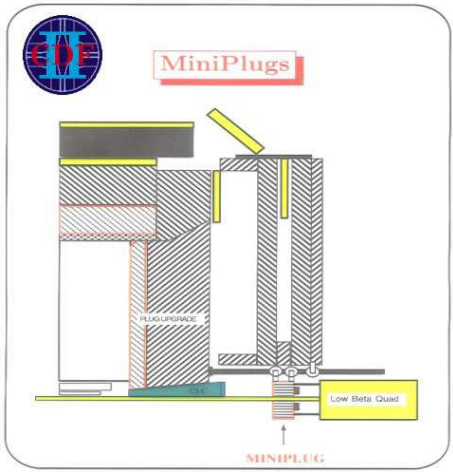,height=2.5in}}
\vspace{-1.cm}
\caption{This is the Miniplug of CDF}
\label{fig:cdfminiplug}
\end{figure}	              
	\item{Add new scintillating counters at large rapidity}
\end{enumerate}
\subsection{D\O\ }
\begin{figure}[htb]
\centerline{\psfig{figure=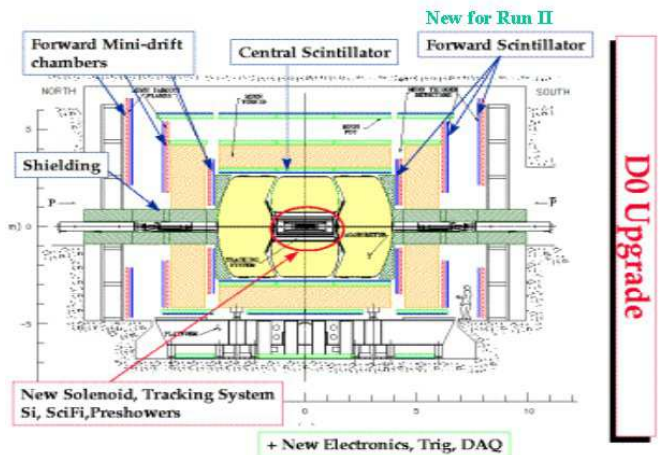,height=2.5in}}
\vspace{-1.cm}
\caption{This figure shows the upgrade of the D\O\ .}
\label{fig:upgrade}
\end{figure}

	D\O\ detector also has been submitted to many modifications and the
figure~\ref{fig:upgrade} summarizes the main features.
\par	
\par
	Now we will describe the Forward Proton Detector for D\O\ .

	\section{Forward Proton Detector}

	The idea of the Forward Proton Detector is to cover experimentally a
large number of topics which will be very important for the progress of the 
diffractive physics. Many details of FPD we can read in the 
reference~\cite{proposal}.
\par
\begin{figure*}[htb]
\centerline{\psfig{figure=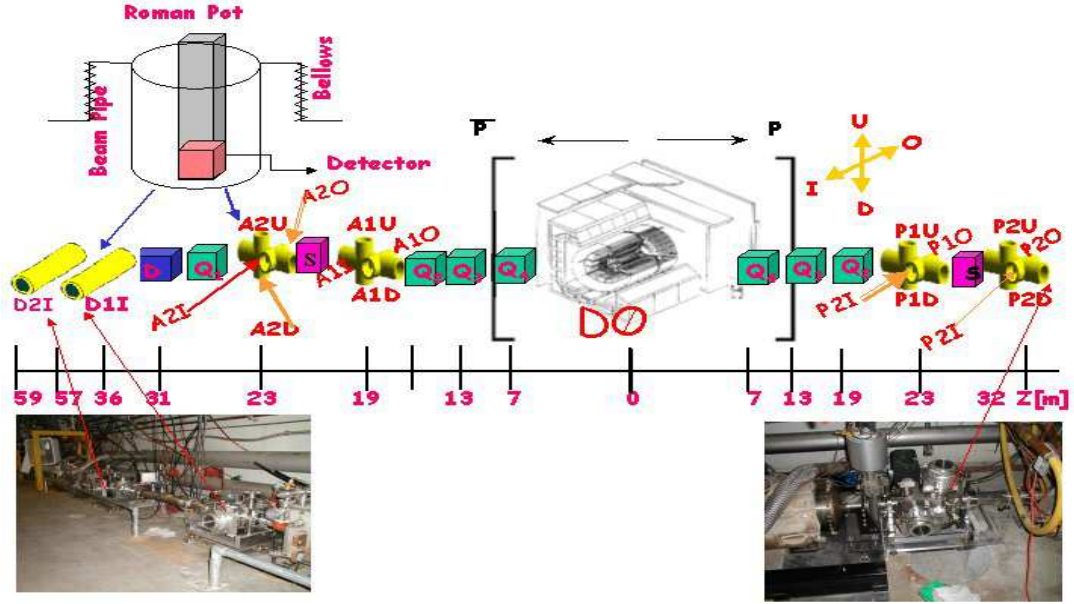,height=3.2in,width=5.7in}}
\vspace{-1.cm}
\caption{This figure shows the FPD in the beam line of the Tevatron, in both
sides of the  D\O\ detector.}
\label{fig:capa}
\end{figure*}

	The Forward Proton Detector consists of 18 Roman Pots arranged on both 
sides of the D\O\ detector as is shown in figure~\ref{fig:capa} where the Roman 
Pots are in the beam line of the Tevatron. We have two castles on the proton 
side indicated by $P1$ \ and  \ $P2$ as is shown in figure~\ref{fig:capa}. The 
orientation is indicated by the additional letter U for up position, D for down 
position, I for inside position and O for outside position of the pots ($P1U, \ 
P1D, \ P1I, \ P1O$, same notation for $P2$). On the anti-proton side we have 
two similar castles labeled by $A1$ \ and \ $A2$ followed by the indication 
of the position similar for the proton side. Two others half castles on the 
anti-proton side are the dipole magnet labeled $D1$ \ and \ $D2$. The approximated 
distances of the pots with respect to the interaction point (indicated by 0 on the 
scale) are shown.
\par	
\par

\subsection{Some Physics Topics for the Run II at Tevatron}

	With the FPD we intend to cover a good phase space region, in order to 
revisit some old and interesting results and create the conditions to
observe new physics topics in diffraction. The combination of the Dipole and 
Quadrupole stations will allow us to get a good sample of data. On 
table~\ref{table:1} we show a comparison between the available data and the 
possible sample to get with FPD.
\par
\begin{table}[ht]
\caption{This table shows a comparison between 
what we have in the present and what is our 
expectation using FPD at the Run II of the 
Tevatron.} 
\label{table:1}
\par
\begin{tabular}{||c |c |c||} \hline \hline  
Experiment    & Dijet Events  &  $E_{T}$ [GeV]  \\ \hline \hline 
     UA8         &  100       &    8            \\ \hline      
     HERA        &  Hundreds  &    5            \\ \hline          
     CDF         &  Thousands &    10           \\ \hline    
                 &  500,000   &    15           \\ \cline{2-3}  
   D\O\ /FPD     &  150,000   &    20           \\ \cline{2-3} 
                 &   15,000   &    30           \\ \hline \hline    
\end{tabular}  
\end{table}
\par
\begin{figure*}[htb]
\centerline{\psfig{figure=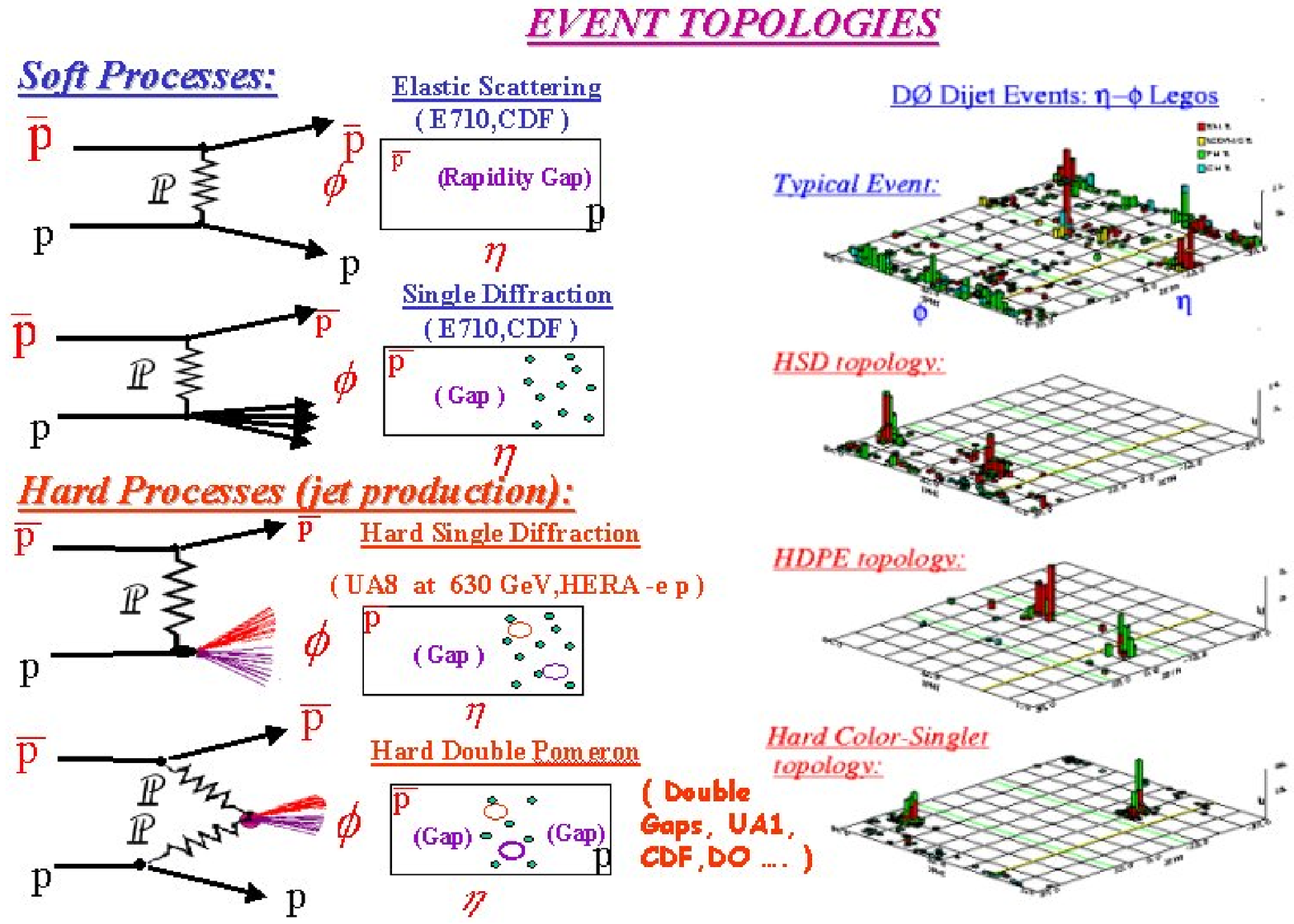,height=4.in,width=5.2in}}
\vspace{-1.cm}
\caption{This figure shows the possible topologies to be studied with FPD in
 D\O\ \ detector. For each topology we have a corresponding lego plot in 
 pseudo-rapidity versus azimuthal angle. We show also the lego plots 
 corresponding to three topologies: the hard single diffraction (HSD), 
 the hard double Pomeron Exchange (HDPE) and the Hard Color-Singlet extracted 
 from D\O\ dijet events.}
\label{fig:topologies}
\end{figure*}

	The topics of physics can be distributed by some topologies. 
In figure~\ref{fig:topologies} we show several possibilities.
\par	
\par
	Some of the topics that can be studied we have:

	\subsubsection{\bf Low and High $ |t| $ Elastic Scattering}
\par 
		The acceptance of FPD spectrometer for low and high $t$ will allow
us to extract elastic scattering events for both physical region. The measurements
of elastic cross sections give a direct connection to the total cross section via
Optical Theorem. It is important to know the elastic slope of the differential
cross section.  The value of the slope for general differential cross sections 
characterizes a specific process which can be associated to a particular 
production (e.g. resonance production).  We can see in figure~\ref{fig:topologies} 
the association of the expected lego plot of pseudo-rapidity versus azimuthal angle 
showing a gap between the scattered proton and anti-proton scattered.
\par
	\subsubsection{\bf Total cross section}
\par
	The results from Tevatron~\cite{E811} experiments are not compatible 
between them if we extrapolate to higher energies. It would be very important 
to have another measurement at these 
energies. It is important to know the behavior of the cross sections with energy
since we have a behavior predicted by the Froissard bound. After the Tevatron only 
the LHC will offer a new opportunity to make these measurements. The measurements 
made at the Tevatron will guide the future measurements at higher energies.
\par
	\subsubsection{\bf Inclusive single diffraction}
\par
	The inclusive single diffraction has many subjects  associated with it.
Particularly for the Tevatron detectors, the Diffractive  mass available, for
single diffraction events, $M_{x} \, = \, 450 $ GeV, makes the extraction 
of heavy flavor physics comfortable. Inclusive single diffraction has been a 
good laboratory for several problems in diffraction physics.  We intend to 
use it to study  jets and to calculate ratios between cross sections of 
different processes.

	\subsubsection{\bf Diffractive Jet production}
\par
	Jets have been largely studied by QCD. The discovered of diffractively 
produced jets~\cite{UA8} by UA8 collaboration was very important for diffractive
physics. This was the main starting point for hard diffraction. It is expected 
that single and double jet diffractive production be exhaustively studied in the 
near future with FPD, making a distinction to the jets produced by the flux of 
color interactions. As we see in table~\ref{table:1} it is possible to get dijet
events with transverse energy $E_T$ higher in run II than in run I.
\par
	\subsubsection{\bf Hard Double Pomeron Exchange}
\par
	Due to the interesting topology, exciting physics topics, double Pomeron
exchange has been largely discussed as the process for many different types of 
production~\cite{levin}. An advantage of the large Diffractive mass produced at 
the Tevatron, in this case $M_{x} \, \simeq \, 100 $ GeV, is the possibility 
to study by direct observation the Pomeron $\times$ Pomeron interactions and the 
associated physics. The instrumentation proposed by FPD/ D\O\ is appropriated to 
face the challenges of the double  Pomeron mechanism, and to produce several 
objects 
not yet observed. Figure~\ref{fig:topologies} shows the double Pomeron exchange
graph and the gaps on its corresponding lego plot. Other topics like Glueballs,
Centauros and Higgs can be exploited also in this topology.
\par 
	\subsubsection{\bf Diffractive Heavy Flavor Production}
\par
	Heavy Flavor physics has been extensively studied in high $p_t$ physics. 
Because experimental results are rare for heavy flavor diffractively produced, there
was not enough attention to this physics. For lack of adequate instrumentation
we have cross sections without a trigger separation for diffractive and 
non-diffractive events. Following the flavour of quarks we can split the studies of
diffractive heavy flavor production, (i) the Charm, (ii) the Bottom and the (iii)
Top Physics. Each one has some particularities to be taken into account in the 
diffractive production~\cite{berger}.
\par
	Heavy flavor physics has been considered almost as a high $p_t$ physics 
only. Diffractive heavy flavor production has not been sufficiently studied. This 
is due mainly to the absence of experimental results in this area as a consequence
of lack of adequate instrumentation to observe diffractive production of heavy 
flavor. 
\par
	\subsubsection{\bf Diffractive W/Z boson production}
\par
	The present results from diffractively produced W and Z bosons are not 
satisfactory, we need more statistics for these events. This is important to better
understand these processes, in order to have a comparison with the current
boson production. Both CDF~\cite{cdf} and  D\O\ have made progress, and the current
results are motivating both collaborations to proceed with new measurements. 
\par
	\subsubsection{\bf Diffractive Structure Functions}
\par
	The study of Diffractive structure functions at the Tevatron would allow 
a comparison with the existing Hera results. To understand the structure of the 
Pomeron one must know its structure function. This type of study has to be 
pursued exhaustively to get better accurate events and to have a clear
interpretation of the Pomeron. One should know how important is the gluon and the 
quark component of the Pomeron. The universality of the Pomeron components is 
important in order to have more clear idea of this object. Practically all topics 
depend of the knowledge of the structure function of the Pomeron. Many 
theoretical developments depend on the experimental results for diffractive 
structure function. It is important to build a good sample of events to study the
diffractive structure functions in $\overline{p} p$ at the energy of the Tevatron.
\par	
	\subsubsection{\bf Glueballs, Centauros and Higgs}
\par
	Since the origin of QCD, Glueballs has been studied  by theoreticians 
and experimentalists.  However, we do not have a significant progress in this
subject. We need more experiments dedicated to the discover of glueballs 
without ambiguity with quark anti-quark competitive states. The family of 
glueballs is big.  Table~\ref{table:2} shows the glueballs (oddballs are also 
shown). Oddballs should have the priority to be examined due to the fact that  
they do not have competition with current $q \bar q$ states, mesons, and the 
$qqq$ states, (baryons)  with the same quantum numbers. It is difficult to 
separate the common hadrons from the glueballs when they appear in the same 
physical region. Glueballs can be produced also in double Pomeron Exchange topology.
We believe that glueballs are largely produced by this mechanism.
\par
\begin{table*}[ht]
\begin{center}
\caption{This table shows possible glueball state configurations with the 
mass and the quantum numbers for each one.} 
\label{table:2}
\par
\begin{tabular}{||c |c |c |c |c |c |c |c||} \hline \hline
\multicolumn{8}{||c||}{Glueballs and Oddballs} \\ \hline
          &     &      &   &     &  \multicolumn{3}{c||} {}           \\ 
 $J^{PC}$ &$(q \bar q)$&2g &3g&ODD& \multicolumn{3}{c||} {MASS (GeV)}\\ 
         \cline{6-8}
          &&&&&\cite{kaidalov} & \cite{morningstar} &\cite{Teper} \\ 
         \hline \hline
$0^{++}$ &YES  &YES   &YES&NO&$1.58$&$ 1.73\pm 0.13$&$1.74\pm 0.05$ \\ \hline         
$0^{+-}$ &NO&NO &YES&YES  &      &               &        \\ \hline     
$0^{-+}$ &YES  &YES   &YES&NO&      &               &        \\ \hline   
$0^{--}$ &NO&NO &YES&YES  &$2.56$&$2.59\pm 0.17$ &$2.37\pm0.27$ \\
         \hline 
$1^{++}$ &YES  &YES   &YES&NO&      &               &        \\ \hline 
$1^{+-}$ &YES  &NO &YES&NO&      &               &        \\ \hline 
$1^{-+}$ &NO&YES   &YES&YES  &      &               &        \\ \hline   
$1^{--}$ &YES  &NO &YES&NO&$3.49$&$3.85\pm0.24$  &        \\ \hline 
$2^{++}$ &YES  &YES   &YES&NO&$2.59$&$2.40\pm0.15$  &$2.47\pm 0.08$\\ 
         \hline 
$2^{+-}$ &NO&NO &YES&YES  &      &               &        \\ \hline 
$2^{-+}$ &YES  &YES   &YES&NO&$3.03$&$3.1\pm 0.18$  &$3.37\pm 0.31$ \\ 
         \hline 
$2^{--}$ &YES  &NO &YES&NO&$3.71$&$3.93\pm 0.23$ &        \\ \hline 
$3^{++}$ &YES  &YES   &YES&NO&$3.58$&$3.69\pm 0.22$ &$4.3\pm 0.34$ \\ 
         \hline 
$3^{+-}$ &YES  &NO &YES&NO&      &               &        \\ \hline 
$3^{-+}$ &NO&YES   &YES&YES  &      &               &        \\ \hline 
$3^{--}$ &YES  &NO &YES&NO&$4.03$&$4.13\pm 0.29$ &        \\ \hline 
          \hline 
\end{tabular}
\end{center}
\end{table*}
\par
\par
	Another topic is the production of  Centauros, which were never 
observed in accelerator particle physics. These objects were discovered in 
Cosmic Ray Physics as events with several unusual characteristics, like the 
production of large multiplicity charged particles accompanied by very few 
photons. For example, as many as 100 charged particles and no more than 3 
$\pi^{0}$~\cite{halzen}. We have enough energy at the Tevatron to produce 
Centauros. Since our diffractive mass is significantly high, we can produce 
them diffractively. The good calorimetry of the D\O\ detector can be very 
useful to observe the absence of electromagnetic activity. 
\par
	Higgs is one of the exciting subject in the next Run II. It is not excluded
that they are produced also diffractively. We have two recently studies giving by
references~\cite{levin}, showing the possibility of Higgs production by double 
Pomeron mechanism.
\par
	\subsubsection{\bf Correlations between $\eta$, t, $M_x$, b, $\xi$, x, $E_T$,..}	
\par
	During the study of any subject in diffractive physics it
would be important to know the behavior of the data with respect to a few number
of kinematics variables and their correlation among themselves.
The kinematical variables are: the pseudo-rapidity $\eta$ defined by 
\[ \eta \, = 
\, -\, \ln{(\tan{\frac{\theta}{2}})},  
\] 
$t$  is the transferred momentum between 
the proton beam and the scattered proton defined by $ t \, = {(P_{Beam} \, - 
P_{Scattered})}^2$; $ M_x \, = \, \sqrt{\xi} \, \sqrt{s} $ is the diffractive mass 
which is 450. GeV for single diffraction and 100. GeV for double Pomeron 
exchange at the energies of the Tevatron ($\sqrt{2}$ TeV); $b$ is the measured 
slope of the differential cross section, which can be selected globally or for a 
particular region of the invariant mass produced diffractively  ( $ \frac{d \sigma}
{dt} \, \propto \, e^{-b(M_x) t} \, ) $; $ \xi \, = \, 1 - x_p \, = \, 
\frac{\Delta P}{P} $ is the fraction of the momentum of the proton carried by the 
Pomeron; $x_p \, \geq \, 0.95 $ is the fraction of the momentum of the proton 
carried by the scattered proton; and $E_T$ = is the transverse energy of the  jet 
produced by hard diffraction; $\theta$ is the polar angle of the scattering and 
$\phi$, the corresponding azimuthal angle.  The distributions of these kinematics 
variables are a phenomenological source of investigation of the possible hidden 
dynamics of the diffraction. The lego plot $\eta$ the pseudo-rapidity, versus 
$\phi$ it is an example of this type of study. The rapidity gap still is a good 
signature for diffractive events. 
\par	
\subsection{Roman Pots}
\par
\begin{figure}[htb]
\vspace{9pt}
\centerline{\psfig{figure=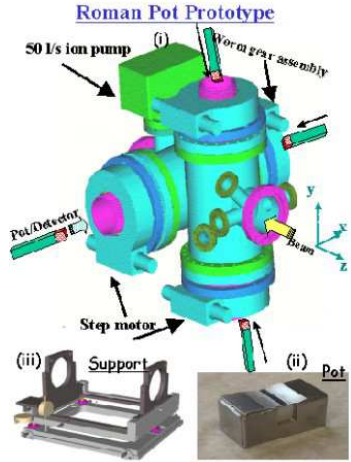,height=2.5in}}
\caption{This figure shows the (i) castle, the (ii) support of the castle and the
 pot itself. The figure shows all main parts of the castle, four directions; 
 in each one a pot is being inserted. The pots moving to be approximated of 
 the beam using step motors. These three parts constitute the mechanical part 
 of the Roman Pots Stations. }
\label{fig:potes}
\end{figure}

	The design of the Roman Pots of D\O\ FPD is shown in figure~\ref{fig:potes}.
They were built by LNLS/Brazil (Laborat\'orio Nacional de Luz Synchroton) as part 
of our regional collaboration ( LAFEX/CBPF -Centro Brasileiro de Pesquisas 
F\'\i sicas;
UFBA -Universidade Federal da Bahia; UFRJ -Universidade Federal do Rio de Janeiro; 
UERJ -Universidade Estadual do Rio de Janeiro; IFT/UNESP -Universidade Estadual 
Paulista; and LNLS). During two years we studied many options for the castle and 
detectors. The castle was made using 316L  steel following the technical 
specifications
to obtain the high vacuum of the Tevatron. The combination of the four views,
quadrupole stations as is shown in figure~\ref{fig:potes} and the dipole stations 
give the possibility to cover more phase space and to get a better acceptance. 
In figure~\ref{fig:potes} we show the castle, the support of the castle and the pot 
itself made by NIKHEF laboratory. In order to get the best pot performance the 
design
decision of the pot window of the pot was submitted to the finite element analysis. 
150 microns foil with elliptical cutout gave the best result. 
\par 
\par	

\subsection{The Detectors}
\par
\begin{figure}[htb]
\vspace{9pt}
\centerline{\psfig{figure=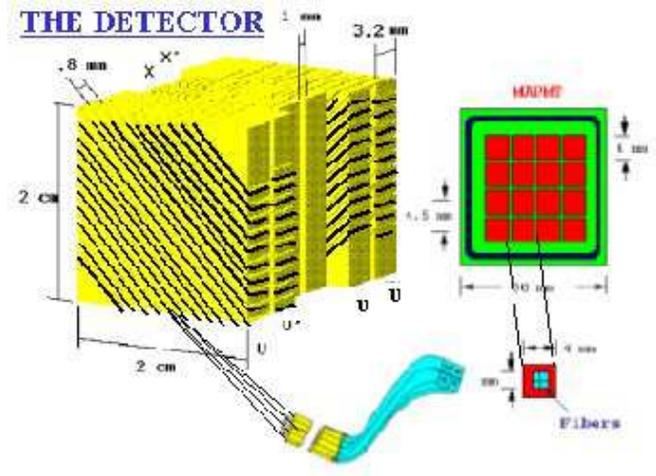,height=2.5in}}
\caption{This figure shows the six planes (u u', v v', x x') of the detectors 
of the FPD spectrometer and the MAPMT.  Each form scintillator fibers is 
connected in the figure join to a cell of the MAPMT.}
\label{fig:detectors}
\end{figure}

	For the detectors we choose the 800 microns scintillator fibers as is shown 
in figure~\ref{fig:detectors}. In this figure~\ref{fig:detectors} we see the frame
of the scintillating fibers and 6 planes X X', U U' and V V' which compose one unity
or one detector to be inserted in the pots. The scintillating fibers are glued to 
clear fibers which guide the signal up to the multi-anode photomultipliers (MAPMT
H6568 from Hamamatsu). We have 16 channels per plane X X' and 20 channels/plane 
U U', V V', giving a total of 112 channels per detector and 2016 channels in total.
Studies about the signals, efficiency and resolution have been made. Scintillating 
fibers are the best option for our detectors among many other possible technologies.
The frame is made of ordinary plastic. The theoretical resolution is 80 microns. 
The acceptance of our detectors has been studied in several views. 
\par 
\par
\begin{figure}[htb]
\vspace{1pt}
\centerline{\psfig{figure=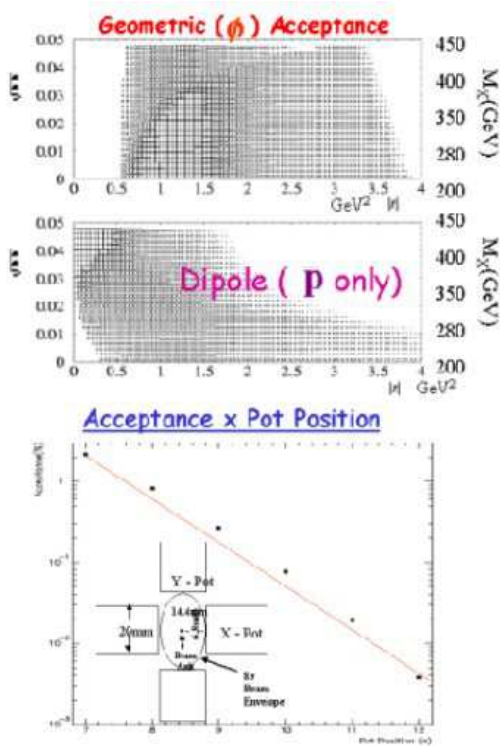,height=4.in}}
\caption{This figure shows the acceptance $\xi$ t for both dipole and
quadrupole stations.  The figure shows also the acceptance versus the
pot position.  We have an idea of the acceptance variation moving the 
pot to be as near as possible of the beans.}
\label{fig:acceptances}
\end{figure}

	
	The geometrical acceptance and the pot position acceptance is given in 
figure~\ref{fig:acceptances}.

\par	

\section{Conclusion}
	The diffractive physics at Tevatron in the next Run II will be very 
exciting.
Practically all possible subjects can be studied with the data obtained by 
CDF and D\O\ FPD. We have shown how interesting are the upgrades including 
Tevatron, Detectors and subdetectors for diffractive physics. 
\par
	The main goal of the experiments in diffractive physics is the hard diffraction.
Many theoretical and phenomenological progress has been done without a corresponding
set of data allowing the comparisons.  New data is urgently needed to
establish limits and to guide our imagination.
\par
	The primary physics goal of the FPD is to measure hard diffraction,
obtaining new data, for the progress of the diffractive physics.  This is an 
opportunity to clarify many hypothesis made theoretically as well as to test the 
present models for diffraction. The data will also give the possibility to 
have a good comparison with the Hera diffractive physics production of last 
years in order to test the universality of the Pomeron. It will also be possible 
to use the FPD to reduce uncertainties on the luminosity for all D\O\ physics 
processes. The double Pomeron
exchange and the physics that can be done with this topology is one of the very 
important subject of this next Run II of the Tevatron. It is not excluded as we 
called attention, that Centauros and Higgs~\cite{levin} can be produced 
diffractively in the double Pomeron topology. 
\par
	We will obtain results which will improve old measurements at lower 
energies and, in some cases, can contribute for clarifying existing results as 
is the case of the total cross sections in Tevatron energy. We gave a list 
of possible topics to be investigated in both hard and soft diffraction. 
\par
	Our schedule to start the data acquisition is the same of the D\O\ 
Detector i.e., at the beginning of the year 2001.  Finally the diffractive physics 
results of D\O\/FPD, will be very important for future projects
at LHC, since it is expected that Diffractive Physics becomes more important when
the energy increases.
\par
I would like to thank the organizing committee of the International Workshop on 
Diffraction in High-Energy Physics, Diffraction 2000, and in particular Prof. 
R.~Fiore, for having invited me to give this talk, giving me the 
opportunity to participate to this interesting Workshop.
All the work behind this paper was done by our colleagues from CDF and D\O\ 
collaboration. We thank the organizing committee of the Diffraction 2000 and 
FAPERJ for partial financial support. 
\par

\end{document}